\documentstyle[preprint,aps]{revtex}
\begin{document}
\draft
\author{Fang-Pei Chen}
\address{Department of Physics \\
Dalian University of Technology \\
Dalian, 116024 P. R. China \\
E-mail: chenfap@dlut.edu.cn}
\date{\today}
\title{A New Study about the Two Formulations of Conservation Laws for Matter Plus Gravitational Field and Their Experimental Test}
\maketitle

\newcommand{\emtd}{\mbox{${\cal T}^{\mu}_{(M)\nu}$}}
\newcommand{\emtdzero}{\mbox{${\cal T}^{0}_{(M)0}$}}
\newcommand{\empd}{\mbox{${\it t}^{\mu}_{(G)\nu}$}}
\newcommand{\empdi}{\mbox{${\it t}^{\mu}_{(G)i}$}}
\newcommand{\emPd}{\mbox{${\cal T}^{\mu}_{(G)\nu}$}}
\newcommand{\empdzero}{\mbox{${\it t}^{0}_{(G)0}$}}
\newcommand{\emPdzero}{\mbox{${\cal T}^{0}_{(G)0}$}}
\newcommand{\empdzeroi}{\mbox{${\it t}^{i}_{(G)0}$}}
\newcommand{\emtdVal}{\mbox{$ 2\ \frac{{\delta W}_{M}}{{\delta g}_{\mu\alpha}}g_{\nu\alpha}$}}
\newcommand{\emtdDef}{\mbox{$\emtd \stackrel{\rm def}{=} \emtdVal$}}
\newcommand{\empdVal}{\mbox{$ 2\ \frac{{\delta W}_{G}}{{\delta g}_{\mu\alpha}}g_{\nu\alpha}$}}
\newcommand{\emPdDef}{\mbox{$\emPd \stackrel{\rm def}{=} \empdVal$}}
\newcommand{\Vi}{\mbox{$v^{\mu \sigma}_{(G) \nu}$}}
\newcommand{\dVi}{\mbox{$\frac{\partial}{\partial x^\sigma} \Vi$}}
\newcommand{\Vii}{\mbox{$v^{\sigma \mu}_{(G) \nu}$}}
\newcommand{\dVii}{\mbox{$\frac{\partial}{\partial x^\sigma} \Vii$}}

\begin{abstract}
The debate on conservation laws in general relativity eighty years ago is 
reviewed and restudied. The physical meaning of the identities 
$\emtd (x) + \emPd (x) = 0$
is reexamined and new interpretations for gravitational wave are given. 
The conclusions of these studies are distinct from the prevalent views, it can
be demonstrated that gravitational wave does not transmit energy (and
momentum) but only transmits information. An experimental test is offered to 
decide which conservation laws are correct.
\end{abstract}


\section{ The Debate in General Relativity Eighty Years Ago}

In 1914, Einstein obtained the conservation laws for matter plus gravitational
field in the form [1,2]

\begin{equation}
\frac{\partial}{\partial x^\mu} (\emtd + \empd) = 0 \ ,
\end{equation}

$\emtd$ is the energy-momentum tensor density for matter field, Einstein 
called $\empd$
the energy-momentum pseudotensor density of gravitational field.

Lorentz in 1916 [3] and Levi-Civita in 1917 [4] proposed successively to use
\begin{equation}
\emtd (x) + \emPd (x) = 0
\end{equation}

or
\begin{equation}
\frac{\partial}{\partial x^\mu} (\emtd + \emPd) = 0 \ 
\end{equation}

as conservation laws for matter plus gravitational field, their propositions
evoked an important debate [2,5,6] on the correct formulation of conservation
laws in general relativity eighty years ago.

The definition of \emtd\ is \emtdDef\ [7]. 
This definition has been accepted universally in theoretical physics. It is
naturally to define the energy-momentum tensor density for gravitational field
by  \emPdDef\ . 
So Lorentz and Levi-Civita adopted this definition. In the above
definition, $W_{M} = \int {\cal L}_M(x) d^4 x$, 
$\delta W_M = \int \frac{\delta W_M}{\delta g_{\mu\nu}}
\delta g_{\mu\nu} d^4 x$;
$W_G = \int {\cal L}_G(x) d^4 x, \delta W_G =
\int \frac{\delta W_G}{\delta g\mu\nu}d^4 x$;
${\cal L}(x) = {\cal L}_M(x) + {\cal L}_G(x)$
is the Lagrangian density of the whole system, 
${\cal L}_M(x)$ and  ${\cal L}_G(x)$ are the matter 
field part and the pure gravitational field part of ${\cal L}(x)$ respectively. 
In General relativity, 
$\emPd = \frac{c^4}{8 \pi G} \sqrt{-g}  (R^{\mu}_{\nu} - \frac{1}{2} {g}^{\mu}_{\nu} R)$.
Some people think that \emPd\ is a pure geometric quantity and can not be
used as the definition of energy-momentum tensor for gravitational field.
This view is incorrect, because the metric tensor $g_{\mu\nu}$ is both 
geometric quantity and dynamic quantity in the theory of gravitation, so 
is \emPd\ .

Eq. (1) can be derived from the local translational symmetry of the 
gravitational system [8,9]. There exist the relations [9]

\begin{equation}
\empd  =  \empdVal - \dVi \ , \  \dVi  =  - \dVii \
\end{equation}

where \Vi\ is determined by the Lagrangian density  ${\cal L}_G$. 
Eq. (1) can also be 
derived from Einstein field equations [10] or from the covariant generalized
conservation laws ${\cal T}^{\mu}_{(M)\nu;\mu} = 0$ [5]. 
\empd\ obtained from distinct methods are different,
but their difference can always be expressed by the relations: 
$''\!\empd -\  '\!\empd = \partial_{\alpha}u^{\mu\alpha}_{\nu}$, where
$\partial_{\alpha} u^{\mu\alpha}_{\nu} = -\, \partial_{\alpha}u^{\alpha\mu}_{\nu}$
or $u^{\mu\alpha}_{\nu} = -\, u^{\alpha\mu}_{\nu}$.

The quantity \emPd\ is a tensor density, but \empd\ is not.
The conservation law Eq. (3) is covariant, but the conservation law Eq. (1)
is not. The well-known serious difficulties in connection with \empd\ do not
exist for \emPd\ and Eq. (3) is more in line with the spirit of general
relativity [3,4].

The logical rationality for the definition of \emPd\ had been acknowledged by
Einstein [2,5], he also acknowledged that one is not entitled to define
\empd\ as a quantity representing the energy-momentum of gravitational field;
but Einstein doubted about the physical meaning indicated by the relation
in Eq. (2). He said that ``Eq. (2) does not exclude the possibility that
a material system disappears completely, leaving no trace
of its existence. In fact, the total energy in Eq. (2) is zero from the
beginning, and the conservation of this energy value does not guarantee
the persistence of the system in any form'' [5]; so he opposed to choose
\emPd\ as the energy-momentum tensor density for gravitational field. 
Care must be taken to that the reason which Einstein opposed \emPd\ is not 
because of its logical trouble or is not dependent on any experimental 
result; it is only owing to that he thought Eq. (2) being nonsensical.
In the following the identities Eq. (2) is reexamined and a new
explanation is given, moreover an experimental test is offered to decide
which definition of gravitational energy-momentum tensor density and which 
formulation conservation laws are correct. It can be shown that Eq. (2) have
a plentiful physical contents and might be tested by experiments.

\section{Reexamination of the Identities $\emtd (x) + \emPd (x) = 0 $ and 
New Interpretations of Gravitational Wave}

Should Eq. (2) cause inevitably a material system disappear completely?
This is the crux of the problem. We must point out that it is infeasible to
determine solely how a material system changes merely using the 
conservation law of energy alone. Moreover it is impossible to determine
whether this material system disappears completely. The change of a material
system must yet obey other laws, such as the conservation law of baryon
number, the second law of thermodynamic, etc. Therefore, 
$\emtd + \emPd = 0$ does not necessarily give $\emtd = 0$, it only 
implies $\emPd (x) = - \emtd (x)$.

Eq. (2) shows that the energy-momentum tensor of the gravitational 
field must coexist with the energy-momentum tensor of material. Their
sum total is equal to zero invariably and their distributional region 
in space-time is the same. These properties make Eq. (2) have a plentiful 
physical content which we shall show below, the consequences of these 
properties would be verified in future experiments. So Eq. (2) is not
nonsensical from a physical point of view.

It is worth noting that, \empd\ and \emPd\ are not independent, they are
interrelated by Eq. (4); via definition of \emPd\ and Eq. (4), Eq. (3)
can be derived from Eq. (1), and on the other hand via Eq. (4), Eq. (1) can 
also be derived from Eq. (3), so Eq. (1) and Eq. (3) are not independent
either. As a better choice, in this paper we shall follow
Lorentz and Levi-Civita to treat  \emPd\,, but not  \empd\,, as the 
energy-momentum tensor density for gravitational field, and to treat Eq. (3),
but not Eq. (1), as the conservation laws of energy-momentum tensor density;
and we shall treat  \empd\ as a subsidiary quantity, then Eq. (1) represents
a number of subsidiary relations. According to Lorentz and Levi-Civita's
formulation of conservation laws, the Einstein's gravitational equations
$\sqrt{-g}  (R^{\mu}_{\nu} - \frac{1}{2} {g}^{\mu}_{\nu} R) = - \frac{8 \pi G}{c^4} \emtd$
are interpreted both as field equations and as conservation laws [3,4], 
therefore, except using \emPd\ as the energy-momentum tensor density for
gravitational field, all deductions of gravitational field equations remain
unchanged. For instance there was still a singularity at the beginning of the 
present expansion phase of the universe.

The existence of gravitational wave is determined by Einstein's gravitational
field equations, the characteristics of these equations show that the
gravitational wave propagate with speed of light [11].

A great number of people follow Einstein's viewpoint, they believe that 
gravitational wave, like electromagnetic wave, is accompanied by 
radiation of energy. They also believe that this radiation has
been verified by PSR 1913 + 16. These views are incorrect. Now
let us show that gravitational wave could not transmit energy
but could transmit information. The basis for the belief that gravitational
wave radiates energy is using \empd\ and applying the following equation

\begin{equation}
- \frac{\partial}{\partial t} \int_V (\emtdzero + \empdzero) d V =
c \oint_{S} \empdzeroi d S_i \ ,
\end{equation}

which can be derived from Eq. (1). 
The surface $S$ enclosing the volume $V$ is taken in vacuum where $\emtd = 0$. 
The right integral in Eq. (5) is interpreted as the
amount of gravitational energy transferred by the gravitational wave across the
surface $S$ in unit time.

If we treat \emPd\,, but not \empd\,, as the energy-momentum tensor
density for gravitational field, new insights about the gravitational
wave can be gained from Eq. (3) (or Eq. (2)). Using these equations, it
is easy to obtain the identity

\begin{equation}
\frac{\partial}{\partial t} \int_{V} (\emtdzero + \emPdzero) d V
= 0 \ .
\end{equation}

This identity can also be derived from Eq. (5) via Eq. (4); and inversely
Eq. (5) can be derived from Eq. (6) via Eq. (4). Eq. (6) indicates that 
there are no energy
transferred from $V$ to outside via the gravitational wave, $i.e.$ 
gravitational wave does not transmit energy. It must be pointed out
that although $\emtdzero + \emPdzero = 0$, 
the values of $\emtdzero$ and $\emPdzero$ may change, they can
be transformed to each other.

It should emphasize that, although the gravitational wave does not
transmit energy
(and momentum), but it could transmit information. When the gravitational
wave passes through a space point, the gravitational field, 
$i.e.$ the metric field of this point will change from
$g_{\mu\nu}(\vec{r}, t)$ into $g_{\mu\nu}(\vec{r}, t+\Delta t)$ 
in the time interval $(t, t + \Delta t)$, these changes
in $g_{\mu\nu}$ convey the information from the source of gravitational wave. 
Some people guess that information should be closely bound up with energy,
they think that the information without energy must not exist. Their
guess is not correct. The gravitational wave is determined fully by
Einstein field equations [11], which are irrespective how to define
the energy-momentum tensor density for gravitational field. If we
adopt the definition \emPdDef\,
the gravitational wave would transfer information without energy.

When there exists gravitational wave, the space-time metric $g_{\mu\nu}(x)$
must change. This change may be considered as that the metric
undergoes a perturbation by writing
$g_{\mu\nu}(x) = \stackrel{\circ}{g}_{\mu\nu}(x) + h_{\mu\nu}(x)$
[7, 10], $\stackrel{\circ}{g}_{\mu\nu}$ is the background metric and 
$h_{\mu\nu}$ is the perturbation; usually 
$|h_{\mu\nu}| \ll |\stackrel{\circ}{g}_{\mu\nu}|$,
the magnitude of $h_{\mu\nu}(x)$ reflects the intensity of gravitational wave.

The solution for weak-field approximations of Einstein field equations
has been given as
$\varphi^\mu_\nu (\vec{r}, t) = \frac{4 G}{c^4} \int
\frac{{(T^\mu_\nu)}_{ret.} d^3 x'} {|\vec{r} - \vec{r'}|}$ [7],
where $\varphi^{\mu}_{\nu} = h^{\mu}_{\nu} - \frac{1}{2} \delta^{\mu}_{\nu} h$.
From this formula, it is evident that the gravitational wave can be
generated only if the energy-momentum tensor of material source is 
changed in space-time. Let $R_{\mu\nu}(x)$ and $\stackrel{\circ}{R}_{\mu\nu}(x)$
be the Ricci tensors corresponding to $g_{\mu\nu}(x)$ and 
$\stackrel{\circ}{g}_{\mu\nu}(x)$, $R_{\mu\nu}(x)$ and 
$\stackrel{\circ}{R}_{\mu\nu}(x)$ must
obey with Einstein field equation
$(R_{\mu\nu} - \frac{1}{2} g_{\mu\nu} R) = - \frac{8 \pi G}{c^4} T_{(M)\mu\nu}$
and 
$(\stackrel{\circ}{R}_{\mu\nu} - \frac{1}{2} \stackrel{\circ}{g}_{\mu\nu} 
\stackrel{\circ}{R}) = - \frac{8 \pi G}{c^4} \stackrel{\circ}{T}_{(M)\mu\nu}$
respectively. It has been proved that [10] 
$R_{\mu\nu} = \stackrel{\circ}{R}_{\mu\nu} + R^{(1)}_{\mu\nu} (h) + 
R^{(2)}_{\mu\nu} (h) + \cdots$,
where $R^{(1)}_{\mu\nu}$ and $R^{(2)}_{\mu\nu}$ are composed of the 
covariant derivatives of $h_{\mu\nu}$,
$R^{(1)}_{\mu\nu}$ is the linear term and $R^{(2)}_{\mu\nu}$ is in the 
second degree term. Either
of these two equations can be rewritten as 
\begin{equation}
\stackrel{\circ}{R}_{\mu\nu} - \frac{1}{2} \stackrel{\circ}{g}_{\mu\nu}
\stackrel{\circ}{R} = -  \frac{8 \pi G}{c^4} (T_{(M)\mu\nu} + W_{(G)\mu\nu})
\end{equation}
Where
\begin{equation}
W_{(G)\mu\nu} \stackrel{\rm def}{=} \frac{c^4}{8 \pi G} 
\{ (R_{\mu\nu} - \frac{1}{2} g_{\mu\nu} R) - 
(\stackrel{\circ}{R}_{\mu\nu} - \frac{1}{2} \stackrel{\circ}{g}_{\mu\nu}
\stackrel{\circ}{R}) \} = \stackrel{\circ}{T}_{(M)\mu\nu} - T_{(M)\mu\nu}
\end{equation}
It is suggested to interpret $W_{(G)\mu\nu}$ or its average value as the 
energy-momentum tensor for gravitational wave. But in our opinion, 
although $W_{(G)\mu\nu}$ might
be used to express gravitational field's energy-momentum tensor relative
to the background space-time, but it does not indicate the energy and
momentum transmitted by gravitational wave. Since
$W_{(G)\mu\nu} = \stackrel{\circ}{T}_{(M)\mu\nu} - T_{(M)\mu\nu} =
T_{(G)\mu\nu} - \stackrel{\circ}{T}_{(G)\mu\nu}$ 
$(T_{(G)\mu\nu} = \frac{1}{\sqrt{-g}}\emPd$ and
$\stackrel{\circ}{T}_{(G)\mu\nu} = \frac{1}{\sqrt{-\stackrel{\circ}{g}}}
\stackrel{\circ}{\emPd})$ and
$T_{(M)\mu\nu}(x) + T_{(G)\mu\nu}(x) = \stackrel{\circ}{T}_{(M)\mu\nu}(x) + 
\stackrel{\circ}{T}_{(G)\mu\nu}(x) = 0$,
therefore at any point of space-time, the total energy-momentum tensor for
matter plus gravitational field is identically equal to zero when the 
gravitational wave generate.

A question then arises as to how gravitational wave can be detected. Although
the gravitational wave does not transmit energy, the equation of geodesic
deviation remains correct. As a result one can still design the detectors
based on the effect of this equation [10, 12]. For instance, Weber's cylinder
can be used to detect the gravitational wave. From the conventional viewpoint,
gravitational wave can transmit energy to a detector. But in our viewpoint,
when the gravitational wave is received, the oscillation energy of the detector
does not come from the gravitational wave, it comes from the local gravitational
field where the detector resides. Because the detector is made of matter,
let $\emtd$ be its energy-momentum tensor density, $\emtd$ must satisfy Eq. (2).
Therefore $\Delta\emtd = - \Delta\emPd, \Delta\emtd$
represents the increased energy of the detector and $-\Delta\emPd$ represents
the decreased energy of the local gravitational field; $\Delta\emtd$ or
$\Delta\emPd$ can be calculated from Einstein field equation.

\section{An Experimental Test to Decide Which Formulation Is Correct}

The microwave background radiation gives us important information of the 
early epoch of the universe [13], the discovery and theoretical interpretation
of this radiation spurs the cosmology to get ahead. It is believed that there
existed also a vast amount of gravitational waves at the earliest epoch of the
universe immediately after ``big bang'', with the expansion of the universe
there should also exist background gravitational waves at present [14].

The microwave background radiation has the type of spectrum of black
body radiation [13]. Do these background gravitational waves also have the type
of spectrum of black body radiation? We shall study this question at once.
The black body radiation is thermal radiation in equilibrium [15]. Einstein 
advanced a semi-quantum theory of thermal radiation in equilibrium [15,16]
and derived the equation
\begin{equation}
\rho(\nu, T) = \frac{8\pi h \nu^3}{c^3} \frac{1}{e^{h\nu/KT} - 1}
\end{equation}
from his theory. Eq. (9) expresses Plank's law of radiation.

The Semi-quantum theory of thermal radiation in equilibrium advanced by 
Einstein is concise and reflect well the essence of this radiation, the
key of Einstein's theory is to think that the radiation wave transmits
energy and it is quantized [15, 16]. The prevalent theory assumes that
the gravitational wave transmits energy, and the graviton, which
is similar to photon, is the quantum of energy for gravitational wave.
Therefore Eq. (9) must be also applicable for gravitational radiation. It is 
possible that this distribution might deform during the evolution of the
universe, however, the pattern of spectrum must be yet regular.

On the other hand, according to Lorentz and Levi-Civita's definition
of energy-momentum tensor density $\emPd$ for gravitational field
and the conservation law Eq. (2), the gravitational wave does not 
transmit energy, consequently for gravitational
wave the concepts such as thermal radiation in equilibrium, the quantum
of energy and distribution of radiation energy are all meaningless. 
These distinguishing features imply that Eq. (9) does not apply for 
gravitational waves, $i.e.$ the background gravitational waves
do not tally with the black body radiation or its variety. 
The types of spectrum
of background gravitational waves should not have any regularity,
they are the results of random process. Therefore, through the
observations of the spectrum types for background gravitational waves,
it might provide an experimental test to decide whether gravitational wave 
transmits energy, $i.e.$ it might judge which is the correct
definition of energy-momentum tensor for gravitational field, 
$\emPd$ or $\empd$? And which is the correct formulation of conservation
laws for matter plus gravitational field, Eq. (1) or Eq. (3) (and Eq. (2))?
We shall wait and see.

\section{Has the Gravitational Energy Radiation Been Verified?}

The orbital energy loss of the binary pulsar system PSR 1913 + 16 has
been confirmed from the observation of the decrease in its orbital period [17].
This observation has been widely interpreted as verification for energy
radiation of gravitational wave. However based on our analysis given
above, this interpretation is problematic.

The theoretical basis of the so called verification is Eq. (5). When using
Eq. (5) to compute the energy change of PSR 1913 + 16, one always assumes:
1 $ \frac{\partial}{\partial t} \int_{V} (\empdzero) d V = 0 $
and 2 except the orbital kinetic energy and gravitational
interaction energy (which belongs to $\emtd$ according to definition), the other
kinds of matter energy do not change for this binary pulsar. 
Evidently the result of computation can not be very accurate based on these
assumptions.
The ratio of the observed to the predicted damping rate 
$\dot{P}^{obs}_{b}/\dot{P}^{pred}_{b}$
for PSR 1913 + 16 is $1.00032 \pm 0.0035$ [18], the coincidence between
$\dot{P}^{obs}_{b}$ and $\dot{P}^{pred}_{b}$ is overprecise. 
As another example $\dot{P}^{obs}_{b}/\dot{P}^{pred}_{b}$ for PSR 1534 + 12
is about $0.83$ [19], the coincidence is less evident. Taking the approximation
in calculation of $\dot{P}^{pred}_{b}$ into account, one can not rule out 
the possibility that
the coincidence of observation and prediction for the damping rate $\dot{P}_{b}$
of PSR1913 + 16 might be accidental [20, 21].

Even if the above coincidence for PSR 1913 + 16 is true, yet it can
not fully prove that gravitational wave transmits energy. Because Eq. (5) can
be derived from Eq. (6), the same value of $\dot{P}^{pred}_{b}$ 
may also be obtained
from Eq. (6) and $\emPd$ [21]. Since Eq. (6) means that gravitational wave
does not transmit energy, it is inevitable to conclude that the
gravitational energy radiation has not been verified.


\end{document}